# A Directed Threshold - Signature Scheme

Sunder Lal [*] and Manoj Kumar [**]

*Dept of Mathematics, IBS Khandari.Dr. B.R.A.University Agra.*
Sunder_lal2@rediffmail.com.in.
**Dept of Mathematics, HCST, Farah – Mathura, [U. P.] – 281122.*
Yamu_balyan@yahoo.co.in

**Abstract.** Directed signature is the solution of such problems when the signed message contains information sensitive to the signature receiver. Generally, in many application of directed signature, the signer is generally a single person. But when the message is on behalf of an organization, a valid sensitive message may require the approval of several people. Threshold signature schemes are used to solve these problems**.** This paper presents a threshold directed signature scheme.

## 1. Introduction

Physical signature is a natural tool to **authenticate** the communication ,but it is useless in electronic messages; one has to rely on other methods like **digital signature**. Digital signature is a cryptographic tool to solve this problems of authenticity in electronic communications. Basically digital signature has a property that anyone having a copy of the signature can check its validity by using some **public-information**, but no one else can forge the signature on another document. This **self-authentication** property of digital signatures is quite suitable for many uses such as broadcasting of announcements and public key certificates, but it unsuitable for many applications.

In many situations, signed message is sensitive to the signature receiver. Signatures on medical records, tax information and most personal/business transactions are such situations. For these situations the signature on the message should be such as only the signature receiver can verify the signature and can prove the validity of the signature to any third party , whenever necessary.. Such signatures are called **directed signatures** [3,4,13,14].In directed signature scheme, the signature receiver B has full control over the signature verification process. Nobody can check the validity of signature without his cooperation.

In most situations, the signer is generally a single person. However when the message is on behalf of an organization, a valid message may require the approval or consent of several people. In this case, the signature is done by more than one consenting rather than by a single person. A common example of this policy is a large bank transaction, which requires the signature of more than one person. Such a policy could be implemented by having a separate digital signature for every required signer, but this solution increases the effort to verify the message linearly with the number of signer. **Threshold signature** is an answer to this problem.

The (t, n) threshold signature schemes [6,7,8,10,12] are used to solve these problems.  Threshold signatures are closely related to the concept of threshold cryptography, first introduced by Desmedt [7,8]. In 1991 Desmedt and Frankel [6] proposed the first (t, n) threshold digital signature scheme based on the RSA assumption. This paper proposed a **(t , n) threshold directed signature scheme** based on Shamir's  threshold signature scheme[21] and Schnorr's signature scheme[20].

The section-2 presents **some basic tools**. In Section-3 we present a **Threshold Directed Signature Scheme**.  Section-4 discusses the **security of the Scheme**. An **illustration** to the scheme is discussed in section-5**.  Remarks** are in section-6.

## 2. Preliminaries

**2.1.** Throughout this paper we will use the following system setting.

- A prime modulous $p$, where $2^{511} < p < 2^{512}$.
- A prime modulous $q$, where $2^{159} < q < 2^{160}$ and $q$ is a divisor of $p - 1$.
- A number $g$, where $g \equiv k^{(p-1)/q} \mod p$, $k$ is random integer with $1 \leq k \leq p - 1$ such that $g > 1$; (g is a generator of order $q$ in $Zp^*$).
- A collision free one-way hash function $h$ [23].

The parameters $p, q, g$ and $h$ are common to all users. We assume that every user A  chooses a random $x_A \in Zq$ and computes $y_A = g^{x_A} \mod p$. Here $x_A$ is the  private key of  A and $y_A$ is the public key of A. For our purpose ,we use the directed signature scheme based on Schnorr's signature scheme [20] and Shamir's threshold scheme [21]. These basic tools are briefly described below.

### 2.2. Schnorr's signature scheme

In this scheme, the signature of A on message $m$ are given by $(r_A, S_A)$, where,

$$r_A = h(g^{k_A} \mod p, m), \text{ and } S_A = k_A - x_A \cdot r_A \mod p.$$

Here  random $k_A \in Zq$  is private to A . The signature are verified by checking the equality

$$r_A = h(g^{S_A} y^{r_A} \mod p, m).$$

### 2.3. Shamir's threshold scheme

Shamir's $(t, n)$ threshold secret sharing signature scheme is a scheme to distribute a  secret key K into $n$ users in such a way that any $t$ users can cooperate to reconstruct K but a collusion of $t - 1$ or less users reveal nothing about the secret.  Shamir's scheme is based on Lagrange interpolation in a field. To implement it, a polynomial $f$ of degree $t - 1$ is randomly chosen in $Zq$ such that $f(0) = K$.

Each user $i$ is given a public identity $u_i$ and a secret share $f(u_i)$. Now any $t$ out of $n$ shareholders can reconstruct the secret $K = f(0)$, by pooling their shares and using

$$f(0) = \sum_{i=1}^{t} f(u_i) \prod_{j=1, j \neq i}^{t} \frac{-u_j}{u_i - u_j} \mod q$$

Here we assume for simplicity that the authorized subset of $t$ users consists of shareholders $i$ for $i = 1,2,3\ldots t$.

### 2.4. Directed signature scheme

Suppose that user A wants to generate a signature on message $m$ so that only the receiver B can verify the signature and that B can prove the validity of signature to any third party C, whenever necessary. This scheme consists of the following steps.

#### 2.4.1. Signature generation by A to B

(a). A picks at random $K_{a_1}$ and $K_{a_2} \in Zq$ and computes

$$W_B = g^{K_{a_1} - K_{a_2}} \mod p \text{ and } Z_B = y_B^{K_{a_1}} \mod p.$$

(b). Using a one-way hash function $h$, A computes $r_A = h(Z_B, W_B, m)$, and then

$$S_A = K_{a_2} - x_A \cdot r_A \mod q.$$

The arrangement of numbers $\{S_A, W_B, r_A, m\}$ is the signature if A on $m$.

#### 2.4.2. Signature verification by B

(a). B computes $\mu = [g^{S_A} (y_A)^{r_A} W_B] \mod p$ and $Z_B = \mu^{x_B} \mod p$.

(b) B computes $h(Z_B, W_B, m)$ and checks the validity of signature by equality

$$r_A = h(Z_B, W_B, m) \mod q.$$

#### 2.4.3. Proof of validity by B to any third party C

(a) B sends to $\{S_A, W_B, r_A, m, \mu\}$ to C.

(b) C checks if $r_A = h(Z_B, W_B, m) \mod q$.

If this does not hold C stops the process; otherwise goes to the next steps.

(c) B in a zero knowledge fashion proves to C that $\log_\mu Z_B = \log_g y_B$ as follows.

- C chooses random $u, v \in Zp$ computes $w = \mu^u \cdot g^v \mod p$ and sends $w$ to B.
- B chooses random $\alpha \in Zp$ computes $\beta = w \cdot g^\alpha \mod p$ and $\gamma = \beta^{x_B} \mod p$, and sends $\beta, \gamma$ to C.

- C sends *u, v* to B, by which B can verify that $w = \mu^u \cdot g^v \mod p$.

- B sends α to C, by which she can verify that

$$\beta = \mu^u \cdot g^{v+\alpha} \mod p \text{ and } \gamma = Z_B{}^u \, y_B{}^{v+\alpha} \mod p.$$

## 3. Proposed Scheme

In this section, we present a **threshold directed signature scheme**. Let *G* be a group of *n* designated users, out of which any *t* members can generate the signature on a message *m* for a user B. The user B can verify the signature and that B can prove its validity to any third party C, whenever necessary. Nobody can check the validity of the signature without the help of B. We describe a construction of threshold directed signature scheme for this situation as follows.

We assume the existence of a trusted share distribution center (SDC) which determines the group secrets parameters and the secret shares $v_i, i \in G$. Let *H* be any subset of G, containing *t* members. We also assume the existence of a designated combiner *DC* who collects partial signatures from each user in the subgroup *H*. Every shareholder in the group has equal authority with regard to the group secret. The generation of the directed signature requires *t* out of *n* shareholders and interaction with *DC*. This scheme consists of the following steps.

### 3.1.1. Group Secret Key and Secret Shares Generation

(a). SDC selects the group public parameters *p, q, g* and a collision free one way hash function *h* as described in section 2.1. SDC also selects a polynomial

$$f(x) = a_0 + a_1 x + \ldots a_{t-1} x^{t-1} \mod q, \text{ with } a_0 = K = f(0).$$

(b). SDC computes the group public key, $y_G$, as, $y_G = g^{f(0)} \mod p$.

(c) SDC computes a secret shares $v_i$ for each member of the group *G*, as,

$$v_i = f(u_i) \mod q.$$

Here $u_i$ is the public value associated with user *i* in the group.

(d) SDC sends $v_i$ to each user in a secret manner.

### 3.1.2. Signature generation by any *t* users

If any *t* out of *n* members of the group agree to sign a message *m* for user B, they generate the signature using following steps.

(a) Each member *i* randomly selects $K_{i_1}$ and $K_{i_2} \in Zq$ and computes

$$w_i = g^{K_{i_2} - K_{i_1}} \mod p \text{ and } z_i = y_B{}^{K_{i_2}} \mod p.$$

(b) Each member makes $w_i$ openly and $z_i$ secretly available to each member of $H$. Once all $w_i$ and $z_i$ are available, every member computes $Z, W$ and $R$ as

$$W = \prod_{i \in H} w_i \bmod q, \quad Z = \prod_{i \in H} z_i \bmod q, \text{ and } R = h(Z, W, m) \bmod q.$$

(c) Each member $i$ modifies his/her share, as $MS_i = v_i \cdot \prod_{j=1, j \neq i}^{t} \frac{-u_j}{u_i - u_j} \bmod q$.

(d) Each member $i$ uses his/her modified share, $MS_i$ and random integer $K_{i_1}$ to calculate the partial signature $s_i$ as, $s_i = K_{i_1} - MS_i \cdot R \bmod q$.

(e) Each member $i$ sends his/her partial signature to the designated combiner DC who collects the partial signatures and produces $S$, as, $S = \sum_{i=1,}^{t} s_i \bmod q$.

(f) DC sends $\{S, W, R, m\}$ to B as signature of the group $G$ for the message $m$.

### 3.1.3. Signature verification by B

(a). B computes $\mu = [g^S (y_G)^R W] \bmod p$ and $Z = \mu^{x_B} \bmod p$.

(b) B checks the validity of signature by verifying $R = h(Z, W, m) \bmod q$.

### 3.1.4. Proof of validity by B to any third party C

This part of the protocol runs as in section- 2.4.3 and we omit it here.

## 4. Security discussions

we now discuss security aspect of the proposed scheme.

(a). Is it possible to retrieve the group secret key $f(0)$ from the group public key $y_G$? No because this is as difficult as solving discrete logarithm problem.

(b). Can one retrieve the secret shares $v_i, i \in G$, from the public value $u_i$? No because $f$ is a randomly and secretly selected polynomial.

(c). Can one retrieve the secret shares $v_i$, integer $K_{i_1}$ and partial signature $s_i$, from the equation $s_i = K_{i_1} - MS_i \cdot R \bmod q$. ?Here the number of unknown parameters is three. The number of equation is one, so it is computationally infeasible for a forger to collect the secret shares $v_i$, integer $K_{i_1}$ and partial signature $s_i$, $i \in G$.

(d). Can the designated combiner *DC* retrieve the group secret key $f(0)$ or any partial information from the following equation, $S = \sum_{i=1}^{t} s_i \mod q$? This is again computationally infeasible.

(e). Can one impersonate a member $i$, $i \in H$.? A forger may try to impersonate a shareholder $i$, by randomly selecting two integers $K_{i_1}$ and $K_{i_2} \in Z_q$ and broadcasting $w_i$ and $z_i$. But without knowing the secret share $v_i$, it is difficult to generate a valid partial signature $s_i$ to satisfy the verification equation,

$$Z = [g^S (y_G)^R W]^{x_B} \mod p, \text{ where }, S = \sum_{i=1}^{t} s_i \mod q.$$

(f). Can one forge a signature $\{S, W, R, m\}$ by the following equation,

$$\mu = [g^S (y_G)^R W] \mod p.?$$

To compute integer $S$ from here is equivalent to solving the discrete logarithm problem. If any forger randomly selects $S^*$ and sends $\{S^*, W, Z, R, m\}$ to B, the receiver B would computes $\mu^* = [g^{S^*}(y_G)^R W] \mod p$ and $Z^* = \mu^{* x_B} \mod p$ and checks if

$$r_B \stackrel{?}{=} h(Z^*, W_B, m).$$

The receiver B can easily identify that someone has forged the signature.

(g). Can $t$ or more shareholders act in collusion to reconstruct the polynomial $f(x)$.?

According to the following equation,

$$f(x) = \sum_{i=1}^{t} f(u_i) \prod_{j=1, j \neq i}^{t} \frac{x - u_j}{u_i - u_j} \mod q,$$

the secret polynomial $f(x)$ can be reconstructed with the knowledge of any $t$ secret shares $f(u_i)$, $i \in G$.

So if in an organization the shareholders are known to each other, the temptation for $t$ of them to collude could be irresistible. As a result, they would find the secret key of the company, which will be continued to be used. This attack does not weaken the security of our scheme in the sense that the number of users that have to collude in order to forge the signature is not smaller than the threshold value. However it is worth pointing out that $t$ or more users can conspire to compute the system secrets.

## 5. Illustration

The following illustration support our scheme for practical implementation. We assume there are four users in the system. Out of four users A, C, E and F any two users, say, A and F can generate the directed signature on a message $m$ for the user B with secret and public key pair $x_B = 6$, $y_B = 8$ respectively. The following steps illustrate this scheme.

### 3.3.1. Group Secret Key and Secret Shares Generation

Let SDC choose $p = 23$, $q = 11$, $g = 18$ and $f(x) =. 3 + 5x \mod 11$, where $f(0) = 3$ is the group secret key. The public values $u_i$ and corresponding secret shares $v_i$ of users are as follow.

| Users | public value ($u_i$) | secret share ($v_i$) |
|---|---|---|
| A | 9 | 4 |
| C | 12 | 8 |
| E | 14 | 7 |
| F | 16 | 6 |

Now the SDC determines the group secret key as $f(0)$ and computes the group public key, $y_G$, as $y_G = 18^3 \mod 23 = 13$.

### 3.3.2. Signature generation by any $t$ users

If any two users A and F out of four users agree to sign a message $m$ for user B, then the signature generation has the following steps.

(a) The user A randomly selects $K_{a_1} = 2$, $K_{a_2} = 7$ and computes $w_1 = 3$, $z_1 = 12$.

Similarly the user F randomly selects $K_{f_1} = 5$, $K_{f_2} = 9$ and computes $w_4 = 4$, $z_4 = 9$.

(b) Both the users A and F make $(w_1, w_4)$ and $(z_1, z_4)$ publicly available through a broadcast channel. Once all $(w_1, w_4)$ and $(z_1, z_4)$ are available, each user in $H$ computes the product $Z, W$ and $R$ as
$W = 12$, $Z = 16$ and $R = h(16, 12, m) \mod 11 = 5$ (let)

(c) The users A and F compute their modified shares as $MS_A = 6$ and $MS_F = 8$.

(d) The user A uses his/her modified share $MS_A = 6$ and random integer $K_{a_1} = 2$ and calculates his/her partial signature $s_1 = 5$.

(e) The user F uses his/her modified shadow, $MS_F = 8$ and random integer $K_{f_1} = 5$ and calculates his/her the partial signature $s_2 = 9$.

(f) Both the users A and F send their partial signature to *DC* who produces a group signature $S = 3$.

(g) DC sends $\{3, 12, 5, m\}$ to B as signature of the group *G* for the message *m*.

### 3.3.3. Signature verification by B

(a). B computes $\mu = [18^3 \cdot 13^5 \cdot 12] \mod 23 = 3$ and $Z = 16$.

(b) B checks the validity of signature by computing $R = 5$.

### 3.3.4. Proof of validity by B to any third party C

(a) B sends $\{3, 12, 5, m, 3\}$ to C, and C checks that $R = 5$.

If this does not hold C stops the process; otherwise goes to the next steps.

(b) Now B proves to C that $\log_3 16 = \log_{18} 8$ in a zero knowledge fashion by using the following confirmation protocol.

(i). C chooses at random $u = 11, v = 13$ and computes $w = 2$ and sends $w$ to B.

(ii). B chooses at random $\alpha = 17$ and computes $\beta = 16$ and $\gamma = 4$ and sends $\beta, \gamma$ to C.

(iii). C sends $u, v$ to B, by which B can verify that $w = 2$.

(iv). B sends $\alpha$ to C, by which she can verify that $\beta = 16$ and $\gamma = 4$

## 6. Remarks

In this scheme, we have used the ElGamal public key cryptosystem to obtain the construction of threshold directed signature scheme. The security of this cryptosystem is based on the discrete log problem. Only $t - 1$ shadows are not sufficient to obtain the group secret key and they will also get no information about the group secret key, until $t$ individuals act in collusion.

In this scheme, there is a designated combiner *DC* who collects the partial signature of the signer. We should note that there is no secret information associated with the *DC*. Every user can compute his modified share under mod $q$. If $q$ is not prime, then the calculations of the exponents is performed mod $\Phi(q)$, which is not a prime. This implies that Lagrange interpolation for calculating the modified shadows will not work (except when $q = 3$, in which case we are not interested). Consider the situation, when $\prod_{j=1, j \neq i}^{t}(u_i - u_j)$ and $q$ are coprime. In this case there is no way to find out the multiplicative inverse of $\prod_{j=1, j \neq i}^{t}(u_i - u_j)$ mod *q*. There is only possibility of selecting the large prime *q* numbers in order for each person to get around this difficulty

This signature schemes are meaningless to any third party because there is no way for him to prove its validity. The only knowledge of *Z* is not sufficient to prove the validity of signature. Signature receiver also has to perform the confirmation protocol in a zero knowledge fashion to prove the validity of signature.

## References

1. Blakely G.R. (1979). Safeguarding cryptographic keys, *Proc. AFIPS 1979 Nat. Computer conf.,* 48, p.p. 313-317.

2. Blake I.F., Van Oorschot P.C., , and Vanstone S., (1986). Complexity issues for public key cryptography. In J. K. Skwirzynski, editor, *Performance limits in communication, Theory and Practice, NATO ASI Series E:Applied Science* – Vol # 142,p.p. 75 – 97.Kluwer Academic Publishers. Proceedings of the NATO Advanced Study Institute Ciocco, Castelvecchio Pascoli,Tuscany, Italy.

3. Boyar, J., Chaum D., Damgard I. and Pederson T., (1991), Convertible undeniable signatures. *Advances in Cryptology – Crypto,* 90, LNCS # 537,p.p.189-205.

4. Chaum D. (1995). Designated confirmer signatures, *Advances in Cryptology Euro crypt, 94* LNCS # 950,p.p..86-91.

5. Chaum D . (1991). Zero- knowledge undeniable signatures. *Advances in Cryptology –Eurocrypt, 90,* LNCS # 473,p.p..458-464.

6. Desmedt, Y. and Frankel Y.(1991).Shared Generation of Authenticators and Signatures. In *Advances in Cryptology –Crypto -91, Proceedings.* p.p. 457-469. New York: Springer Verlag.

7. Desmedt, Y.(1988). Society and group oriented cryptography. In *Advances in Cryptology –Crypto -87, Proceedings.* p.p. 457-469. New York: Springer Verlag.

8. Desmedt, Y.(1994). Threshold cryptography. *European Transactions on Telecommunications and Related Technologies.*Vol. 5,No. 4, p.p.35 – 43.

9. Diffie W. and Hellman M.. (1976), New directions in Cryptography, *IEEE Trans. Info.Theory.*31.pp. 644 - 654.

10. Gennaro R., Jarecki Hkrawczyk S., and Rabin T. 1996. Robust threshold DSS signature. *Advances in Cryptology – Euro Crypto - 96, Proceedings.* p.p.354 - 371. Berlin-Heidelberg: Springer Verlag.

11. Guillou, L.C. and Quisquater J.J.. (1988), A practical zero-knowledge protocol fitted to security microprocessors minimizing both transmission and memory. *"Advances in Cryptology –Eurocrypt, 88,* LNCS # 330,p.p.123 - 128.

12. Harn L.(1993). (t,n) Threshold signature scheme and digital multisignature. *Workshop on cryptography and Data security, Proceedings.* June 7-9: p.p.61-73. Chung Cheng Institute of Technology, ROC.

13. Lim C.H. and Lee P.J. (1993). Modified Maurer-Yacobi, scheme and its applications. *Advance in cryptology –Auscrypt,* LNCS # 718, p.p. 308 – 323.

14. Lim C.H. and P.J.Lee. (1996). Security Protocol, In Proceedings of International Workshop, (Cambridge, United Kingdom), Springer-Verlag, LNCS # 1189.

15. Mullin R.C., Blake I.F., Fuji – Hara R. and Vanstone S.A.. (1985).Computing Logarithms in a finite field of characteristic two. *SIAM J. Alg.Disc.Math.,* p.p.276 – 285.

16. NIST. (1994), Digital signature standard. *FIPS PUB,* 186.

17. Okamoto T. (1994), Designated confirmer signatures and public key encryption are equivalent. *Advances in Cryptology – Crypto, 94* LNCS # 839, p.p..61-74.

18. Odlyzko. A.M. (1984). Discrete logs in a finite field and their cryptographic significance. In N.Cot T.Beth and I.Ingemarsson,edidors, *Advances in Cryptology – Eurocrypt, 84,* LNCS # 209, p.p..224 - 314.

19. Rabin, T. (1998). A simplified approach to threshold and proactive RSA. In *Advances in Cryptology –Crypto -98, Proceedings.* p.p. 89-104. New York: Springer Verlag.

20. Schnorr C.P. (1994). Efficient signature generation by smart cards, *Journal of Cryptology,* 4(3), p.p.161-174.

21. Shamir A. (1979). How to share a secret, *communications of the ACM*, 22: p.p. 612 - 613.

22. Yen S.M. and Laih C.S. (1993). New digital signature scheme Based on Discrete Logarithm, *Electronic letters,* Vol. 29 No. 12 pp. 1120-1121.

23. Zheng, Y., Matsummoto T. and Imai H. (1990). Structural properties of one – way hash functions. *Advances in Cryptology – Crypto, 90,* Proceedings, p.p. 285 – 302, Springer Verlag.

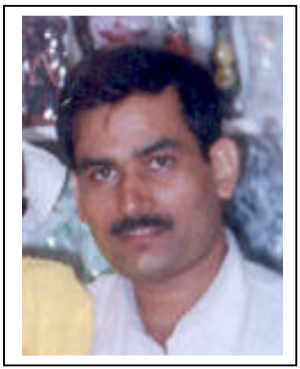

**Manoj Kumar** received the B.Sc. degree in mathematics from Meerut University Meerut, in 1993; the M. Sc. in Mathematics (Goldmedalist) from C.C.S.University Meerut, in 1995; the M.Phil. (Goldmedalist) in *Cryptography*, from Dr. B.R.A. University Agra, in 1996; submitted the Ph.D. thesis in *Cryptography*, in 2003. He also taught applied Mathematics at DAV College, Muzaffarnagar, India from Sep, 1999 to March, 2001; at S.D. College of Engineering & Technology, Muzaffarnagar, and U.P., India from March, 2001 to Nov, 2001; at Hindustan College of Science & Technology, Farah, Mathura, continue since Nov, 2001. He also qualified the *National Eligibility Test* (NET), conducted by *Council of Scientific and Industrial Research* (CSIR), New Delhi- India, in 2000. He is a member of Indian Mathematical Society, Indian Society of Mathematics and Mathematical Science, Ramanujan Mathematical society, and Cryptography Research Society of India. His current research interests include Cryptography, Numerical analysis, Pure and Applied Mathematics.